\documentstyle{article}

\begin{document}
\large
\oddsidemargin 0.55cm 
\vskip 1.5cm
\title{\centerline{NEGATIVE MAGNETORESISTANCE}
\centerline{IN THE NEAREST-NEIGHBOUR HOPPING CONDUCTION}
\centerline{IN GRANULAR GOLD FILM}
}
\vspace{5pt}
\author{\leftline{{\Large B. I. Belevtsev$^{*}$, E. Yu. Beliayev, Yu. F. Komnik,  
E. Yu. Kopeichenko}}\\
\leftline{B. Verkin Institute for Low Temperature Physics \& Engineering}\\
\leftline{Nat. Acad. of Sci. of Ukraine, Lenin Prosp. 47, 310164, Kharkov,
Ukraine}}
\date{}
\maketitle
\par
\vspace{5pt}
\leftline{ABSTRACT} 
\par
The low temperature (0.5--55 K) conduction of semicontinuous
gold film vacuum deposited at $T \approx 50$~K is studied. The film is 
near the percolation threshold (thickness 3.25 nm). Its resistance 
is extremely sensitive to the applied voltage $U$. At low enough $U$ 
the film behaves as an insulator (two-dimensional granular metal). 
In this state the dependences $R(T) \propto \exp (1/T)$ (for $T \leq 20$~K) 
and $R(U) \propto \exp (1/U)$) (for $T \leq 1$~K and $U > 0.1$~V) are 
observed. Magnetoresistance (MR) is negative and can be described 
by $\Delta R(H)/R(0) \propto -H^2/T$. This negative MR which manifests itself 
for nearest-neighbour hopping is rather uncommon and, up to now, has not been 
clarified. The possible mechanisms of such case of negative MR are discussed.
\vspace{5mm}
\par
\leftline{1. INTRODUCTION}
For the last 20 years the significant progress has been achieved in
understanding of transport properties of electrons in granular metals
(GM) [1-4]. In general case GM represents a random mixture of metal
and insulator, i.e. percolating system. In this connection the
conduction of GMs is determined in an essential degree by the ratio of the
metal volume fraction $p$ and the critical fraction $p_c$ (percolation
threshold). Three regimes of electron conduction are distinguished
\cite {Abel75, Sheng92}: (1) Metallic regime ($p > p_c$). 
(2) Dielectric regime ($p < p_c$). (3) Transition regime ($p \simeq
p_c$). In the dielectric regime the conduction of GMs is by a hopping
mechanism in which the charge carriers are transported from one metal
grain to another via thermally-activated tunnelling. It is assumed \cite
{Sheng92}, that in GMs both the nearest-neighbour hopping (NNH) and the
variable-range hopping (VRH) \cite {Mott79, Shklov84} are possible.
\par 
Because of the activated tunnelling of electrons between grains,
the GMs are not classical percolation systems.  Sufficiently strong
external influence on the probability of activated tunnelling can result
in essential change of percolating and conducting properties of GMs. 
Such 
\newline
----------------------------------- \newline
$^{*}$ Corresponding author
\newline
 influence can be exerted by an applied electrical field $F$. For
example, in \cite {Bel86, Bel88} the pronounced variations of conducting
properties of In and Bi granular films under the influence of changing in
$F$ was found. The results of these investigations can be considered as
the evidence of strong effect of an electrical field on the degree of
electron localization in grains. It gives the reason to expect under
certain conditions the occurrence of metal-insulator transition (MIT) in
GMs under the change of $F$. In the present work we report on detection
of such kind of transition in discontinuous gold film near the
percolation threshold.{\LARGE{\footnote {\normalsize {For two-dimensional
systems $p$ is the metallic surface coverage.}}}} It may be assumed that in 
dielectric regime the external magnetic field must also exert noticeable 
influence on the probability of activated tunneling between grains. In 
this connection it should be point out the very interesting results of work 
\cite {Pakhom} in which the negative magnetoresistance (MR) in the 
NNH conduction for the bulk GM was found.  Such  case of negative MR for 
insulators is unusual, and its nature has not been clarified till now.
\par
Ultra thin discontinuous gold films near the percolation threshold were 
used as the objects of research. The films with sheet resistance $R_{\Box}$ 
about 10~k$\Omega$ were prepared by vacuum deposition onto a cold 
substrate ($\simeq 50$~K). The conduction of films with the above-mentioned 
values of $R_{\Box}$ has appeared to be rather sensitive to the applied 
voltage $U$.  At low enough temperatures, decreasing the magnitude of $U$, 
we had the opportunity to transfer films from metallic to dielectric
regime of conduction. Such transition, besides the  essential change 
of temperature dependences of resistance, is  accompanied  by the change 
in the sign  of the MR (from positive to negative).  The observed 
transition from the dielectric to the metallic regime of conduction with 
increase in $U$ is caused by the influence of electrical field on the 
probability of electron tunnelling between metal grains (percolation 
clusters). Same as in work \cite {Pakhom} we found that negative MR of the 
investigated film in the dielectric regime has manifested itself  in the NNH 
conduction. We found however the significant distinction from work 
\cite {Pakhom} which will be discussed below.  The results obtained can be 
used for the development of theoretical models of this interesting phenomenon.
\par 
The main aim of our paper is to represent and discuss the peculiarities of 
observed negative MR in NNH conduction. But for better understanding of 
properties of investigated system we will shortly describe also its 
conduction behavior in dielectric regime and main peculiarities of 
transition from dielectric to metallic state in high enough electrical field. 
\vspace{5mm}
\par
\leftline{2. EXPERIMENTAL PROCEDURES}
The preparation of percolation gold films and the measurement {\it in
situ} their conducting properties were carried out in a high-vacuum
cryostat with $^{3}$He and superconducting solenoid. In the following
the properties of one of the investigated films are described.{\LARGE
{\footnote {\normalsize{The similar behavior of conductivity was
characteristic of other investigated by us gold films, prepared under
conditions described in the present paper.}}}} This film with effective
thickness $L\approx 3.25$~nm (determined with use of the
quartz-crystal monitor) was deposited at the pressure $\approx 10^ {-6}
$~Pa onto a substrate of a single-crystal sapphire with deposition rate
0.015 nm/s at substrate temperature $\simeq 54$~K. The purity of 
initial material was 99.99~\%. Immediately after preparation the
film had resistance $R_{\Box} \simeq 4.87$~k$\Omega$ at $U \simeq
10$~V. For stabilization of the film structural state it was heated 
to $T \simeq 63$~K, then its resistance has decreased to 
$R_{\Box} \simeq 4.5$~k$\Omega$. The film sizes were: $2\times
0.1$~mm.  The conduction of the film was studied by means of recording
of voltage-current characteristics (VCC) at various temperatures (in the
range 0.5--55~K). In the process the source of a stabilized voltage
was used which allowed to change the voltage $U$ in the range from 11 to
0.001~V. In addition to the VCC records, the measurements of MR in a
perpendicular magnetic field $H$ with magnitude up to $\approx 4.5$~T
were made.
\vspace{5mm}
\par
\leftline{3.RESULTS AND DISCUSSION}
\vspace{5mm}
\leftline{{\bf 3.1. Conduction in dielectric regime}}
\vspace{6pt}
The conductivity of the film was essentially non-Ohmic. The temperature 
variations in $R$ at low $U$ and voltage variations in $R$ at low $T$ were 
exponentially great. At the same time at large values of $U$ and $T$ the 
variations in resistance are quite small. At small $U$ ($\leq 0.05$~V) the
temperature dependence $R(T)$ corresponds to simple exponential dependence
\vskip 0pt 
$$ R(T) \propto
\exp \left (\frac{E_{0}}{kT} \right ), \eqno(1)
$$
\vskip 6pt
\noindent where $E_{0} \simeq 1.7$~meV (curve 1 in Fig.~1). As the applied 
voltage $U$ increases the functional dependence (1) remains true in some 
temperature range, but resistance variations are no more exponentially high 
(curves 2 and 3 in Fig.~1). The simple dependence (1) (Ahrrenius law) 
signifies that for investigated film the NNH conduction in dielectric regime 
takes place. At sufficiently low temperatures ($T < 1$~K) identical simple 
exponential dependence $R(U)$ takes place:
\vskip 0pt
$$
R(U) \propto \exp \left (\frac{U_{0}}{U} \right ), \eqno(2)
$$
\vskip 6pt
\noindent where $U_{0} \simeq 2$~V (curve 1 in Fig.~2). 
\par
As the voltage $U$ increases the temperature dependence $R(T)$ becomes
weaker and at large enough voltage $U \approx 10$~V in the range 5-55~K
it corresponds approximately to the relationship $\Delta R (T) \propto
\ln T$. The logarithmic dependence of this kind is consistent with the weak 
localization (WL) and electron-electron-interaction (EEI) effects 
in two-dimensional (2D) systems \cite {Lee85, Alt85}.
\par
The results obtained testify, thus, that the transition from the dielectric 
to the metallic regime of conduction occurs in the
investigated film as the applied voltage increases. As this takes place
the resistance $R_{\Box}$ decreases from $\simeq 10^{7}$~$\Omega$ to
$\simeq 5\cdot 10^{3}$~$\Omega$.  Along with the change of the temperature
dependences of resistance this transition is accompanied by the change of
the nature and the sign of the MR (Fig.~3).  For low-Ohmic states ($R_{\Box} 
\leq 12$ k$\Omega$) at rather high voltages ($\geq 5$~V) the MR was positive.
It is well known \cite {Dump,Gersh95,Bel95a,Bel95b}, that gold films in 
the regime of WL have  positive MR, caused by the strong spin-orbit 
scattering which is characteristic of gold.	We have confirmed that it 
is also true for low-Ohmic states of the film investigated. To do this we 
have compared the experimental MR curves with the appropriate theoretical
expression \cite {Lee85,Alt85} and found a good agreement. At doing so we 
took into account the percolating nature of our film by considering the 
magnitude of sheet resistance $R_{\Box}$ as a fit parameter (s. 
\cite{Gersh95,Buch84}). 
\par
Thus the MR behavior of the investigated film at rather high voltage ($
> 5$~V) is consistent with the WL effects in percolating 2D systems.  As
the voltage is decreased the dependence $R(T)$ became stronger (Fig.~1)
and approached the exponential dependence (1), i.e. the transition in
the dielectric regime of conduction occurred.  This transition is
accompanied by the change in the sign of MR from positive to negative 
(Fig.~3).
\par
The observed $R(T)$ and $R(U)$ dependences [the Eqs. (1) and (2)] are 
in agreement with the 
current concept of GM conduction in the dielectric regime. It is
experimentally established, that the temperature dependences of
conductivity $\sigma$ in this regime have the following form: $\sigma
\propto \exp (-1/T^ {\alpha}) $, where $\alpha$ takes different values 
within the limits 0.25--1.0 \cite {Sheng92}. 
All these dependences have theoretical
explanations [1--4] which are based on the assumption that
the hopping conduction is determined by the joint contribution of two
processes: tunnelling and thermal activation. The case $\alpha = 0.25$,
for example, is attributed to VRH. The simple exponential 
dependence $R(T) \propto \exp (1/T)$~($\alpha = 1$) observed in our 
study corresponds to the NNH.
\par
The observed voltage dependence (2) (Fig.~2) also corresponds well to
the theoretical dependence $\sigma (U)$ for GMs in the dielectric regime
at sufficiently strong electrical fields $F$ and low temperatures \cite
{Abel75,Most89}:
\vskip 0pt
$$
\sigma \propto \exp\left (-\frac{\chi \phi_{i}}{eF} \right ),
\eqno(3)
$$
\vskip 6pt
\noindent  where  
$\chi = \hbar^{-1}(2m\phi_{i})^{1/2}$ ($m$ is the electron mass,
$\phi_{i}$ is the effective barrier height), $\chi^{-1}$ is decay
length of electron wave function in dielectric. The Eq. (3) is correct 
only in the moderate fields ($eFs \ll \phi_{i}$, where $s$ is the 
tunneling distance). In this case the electrical field affects
the form and reduces the effective height and width of a potential
barrier between grains, but electrons remain localized in grains.  In
sufficiently large fields electron is emitted as a free particle in
dielectric layer between grains (direct field ionization) \cite{Most89}.
In the strong-field limit electrons can become nearly free (the
probability of tunnelling is close to unity). In our opinion it is this
effect of electrical field on the tunnelling probability that is
responsible for the observed transition from the dielectric to the
metallic regime of conduction in the gold film as the applied voltage 
is increased.
\par
It follows from our data that not only the increase in applied voltage but 
also the increase in temperature can result in transition in temperature 
dependences of resistance from dielectric to metallic regime of conduction. 
The general (and, we hope, obvious) picture of influence of $T$ and $U$ is 
presented in Figure~4. It can be seen that in such representation the 
equiresistance lines are nearly symmetrized with respect to logarithmic 
scales of $T$ and $U$. The horizontal shading marks the region of positive 
MR (in remaining part the negative MR takes place). 
The vertical shading marks the region where the relation (1) is true. Therefore 
the  negative MR occurs in the region where the Arrhenius law (1) holds (and 
also in adjoining not shaded region, which corresponds to the transition or 
intermediate regime between the dielectric and metallic regimes). 
\par
It can be seen from Fig.~4 that low-Ohmic ($ < 10^{4}$~$\Omega$) states of
investigated film can be attained by ways of the increase of temperature 
(up to $T \geq 20$~K) or the increase of applied voltage ($U \geq 2$~V). In
the first case the transition occurs from activated to non-activated tunnelling 
when the thermal energy $kT$ becomes comparable or larger than the activation
energy $E_{0}$ [see Eq.~(1)]. The mechanism of applied voltage influence on 
the resistance was discussed above.
\par
\vspace{5mm}
\leftline{{\bf 3.2. Magnetoresistance in the nearest-neighbor hopping 
conduction}}
\vspace{6pt}
\par
We have noted above that the MR is negative in dielectric regime. 
At rather low $U$ the MR is described by the following 
square-law dependence: $\Delta R(H)/R = -A(T)H^{2}$ (Fig. 5), where 
$A(T)\propto 1/T^{n}$, $n \approx 1$ (Fig.~6). Therefore the influence of 
a magnetic field on resistance can be given by
\vskip 0pt
$$
\frac{\Delta R_{\Box}(H,T)}{R_{\Box}(0,T)} = -B\frac{H^{2}}{T}, \eqno(4)
$$
\vskip 6pt
\par
The negative MR in the dielectric regime of conduction was revealed 
in our investigation in the NNH conduction. 
This is testified by the temperature dependence of conductivity of the 
investigated film, which is described by the Arrhenius law (Fig.~1), and
the exponential voltage dependence $R(U) \propto (1/U)$  at low 
enough temperatures (Fig.~2).
\par
Previously the negative MR in the NNH conduction had been observed in
\cite{Pakhom} in the three-dimensional GM (composite systems from Al and 
Al$_{2}$O$_{3}$ powders). It was found in this work, that $\Delta
R(H) \propto -H$ (i.e. the linear dependence on $H$), but the
temperature dependence of MR was not determined. For the
investigated percolation gold film we have found the square-law
dependence on $H$ and have determined the temperature dependence (see
Eq.~(4)).{\LARGE
{\footnote {\normalsize{We can not properly compare our results with that 
of \cite{Pakhom} since the system studied in that work has (in contrast to 
our film) very small grain size (diameter is about 3.5 nm) and the 
negative MR was investigated only at high temperatures (at $T = 77$~K 
and 293 K).
}}}} 
\par
For an explanation of negative MR in hopping conduction the interference 
models \cite {Nguen85,Shklov91,Sivan,Botg94} are widely used. This approach
is based on the considering of the interference of 
multiply scattered tunneling paths in hopping probability with taking 
into account the scattering events on the intermediate impurities. 
The consideration of papers \cite{Nguen85,Shklov91,Sivan}, however,
is restricted to the VRH, so that their results are not directly 
applicable to our experimental data. The prediction of model 
\cite{Botg94} ($\Delta R(H)\propto -H^{2}/T$ in conditions of fulfillment 
of the Arrhenius law $R(T) \propto \exp (1/T)$) formally is in complete 
accordance with our results. 
This model is developed, however, for lightly doped semiconductors,
and it remains unclear, how far it is applicable for the GMs.  
\par
We can point out also other two relevant theoretical models \cite{Eto,Wang} 
of negative MR in NNH which could be taken into account at explanation of 
our results. 
In the framework of Hubbard model the following mechanism of negative MR 
in NNH conduction is possible \cite{Eto}. When magnetic field is applied,
the electron spins tend to be parallel to the direction of the field owing 
to the Zeeman effect and hence be parallel to each other. The electron 
correlation effect is less for the electrons having parallel spins. As 
a result, the localization length increases with magnetic field. This 
theory (as was pointed out in \cite{Eto}) can be applied to GM with small 
grains, in which the electron energy level separation is large (about 100 K). 
Such level separation can take place in grains with size of a few nanometers. 
For this reason we beleive that model \cite{Eto} can not be applied to our 
film which is near the percolation threshhold and has rather large grains.{\LARGE
{\footnote {\normalsize{From the MR study of WL effect in metallic regime we 
have evaluated that the phase coherence length $L_{\varphi} \simeq 20$~nm at 
$T \simeq 5$~K. The metallic grain size must be larger than $L_{\varphi}$ 
(this is the necessary condition for observation of WL effect).}}}}
\par
May be more suitable, in our opinion, is the model \cite {Wang} of the 
negative
MR in GM with large metal grains and near the percolation threshold (i.e. 
for the case of small distances between energy levels in grains). The
occurrence of negative MR in this model is supposed to be due to the
influence of the magnetic field on electron orbital motion, with the
result that the change of electron wave functions on the Fermi level
occurs. It can lead to the increase of the overlap of wave functions in
neighbor grains and, therefore, to the increase of hopping probability.
The initial prerequisites for model \cite{Wang} (large grains and the
proximity to the percolation threshold) correspond to the system
investigated in our work. The work \cite {Wang} is, however, only
a ``prelude''. In it only the case $T = 0$ is considered and the form
of  $R(H)$ dependences is not determined.  The nature of the negative MR in
GM in NNH conduction remains, thus, actually unknown. We hope, that the 
experimental results obtained in the present work will promote the 
development of theoretical models of this interesting phenomenon.
\par

\vspace{5mm}
\par
\leftline{CONCLUSION}
The transition from metallic to dielectric regime of conduction has been 
found for discontinuous gold film at decreasing the electrical field, 
applied to the sample. The transition is accompanied by the change of the 
sign of the MR from positive to negative. The negative MR of the film
investigated in the dielectric regime manifests itself for the electron hops 
between the nearest neighbours. The nature of this effect remains unknown.  
The results of the work can be used for the development of theoretical 
models of this interesting phenomenon.
\vspace{5mm}
\par
\leftline{{\bf Acknowledgments}}
\par
The authors are grateful to O. Bleibaum (Technische Universit\"{a}t 
Magdeburg, Germany), P. Sheng and X.R. Wang (The Hong Kong University 
of Science and Technology) for fruitful comments about some aspects of 
the problem of negative MR in granular dielectric systems in the regime 
of nearest-neighbour hopping. In addition we thank X. R. Wang for sending 
us his paper prior to publication. 
\newpage
{
}
\newpage
\centerline{FIGURE CAPTIONS}
\vspace{2mm}
\begin{itemize}
\item[Fig.~1.] $\ln R$ versus $1/T$ at $U = 0.05$~V (1), 0.10 V (2), 
0.2 V (3), 0.5 V (4), 1.0 V (5).
\item[Fig.~2.] $\ln R$ versus $1/U$ at $T = 0.51$~K (1), 2.0 K (2), 
4.0 K (3), 7.0 K (4), 10.0 K (5).
\item[Fig.~3.] The relative variation of resistance in magnetic field at 
$T = 5$~K. The curves correspond to $U = 5$~V (1), 2.5 V (2), 2.0 V (3), 
1.5 V (4), 1.2 V (5), 1.0 V (6), 0.8 V (7), 0.5 V (8).
\item[Fig.~4.] The general picture of influence of $T$ and $U$ on film 
resistance. Bold lines correspond to the resistances 
$R_{\Box} = 1\times 10^{7}$~$\Omega $~(1), $1\times 10^{6}$~$\Omega $~(2), 
$1\times 10^{5}$~$\Omega $~(3), $1\times 10^{4}$~$\Omega $~(4). 
Thin lines (between bold ones) represent the intermediate values of 
resistance $2\times 10^{n}$ and $5\times 10^{n}$ where $n =$~4, 5, or 6.
Upper thin line corresponds to $R_{\Box} = 5\times 10^{3}$~$\Omega$, 
lower one --- to $2\times 10^{7}$~$\Omega$.   
\item[Fig.~5.] The $-\Delta R(H)/R$ as a function of $H^{2}$ 
at $T = 3$~K (1), 4 K (2), 5 K (3), 6 K (4), 7 K (5), 8K (6), 10 K (7),
12 K (8). For each temperature these dependences 
correspond to the minimal value of applied voltage $U$. The lines 
correspond to the relationship  $-\Delta R(H)/R \propto H^{2}$.
\item[Fig.~6.] The dependence $A = f(1/T)$. The function $A(T)$ describes
the temperature dependence of MR of investigated film in accordance with 
the expression $\Delta R/R = - A(T)H^{2}$.
\end{itemize}

\normalsize
\begin{thebibliography}{50}
\bibitem{Abel75}  B. Abeles, P. Sheng, M. D. Coutts, and Y. Arie, 
Adv. in Phys. 24 (1975) 407.
\bibitem{Sheng83} P. Sheng and J. Klafter, Phys. Rev. B27 (1983) 2583.
\bibitem{Adk89} C. J. Adkins, J. Phys.: Cond. Matter 1 (1989) 1253.
\bibitem{Sheng92} P. Sheng, Phil. Mag. B65 (1992) 357.
\bibitem{Mott79}  N. F. Mott and E. A. Davis,  Electron processes 
in non-crystalline materials (Clarendon Press, Oxford ,1979).
\bibitem{Shklov84} B. I. Shklovskii and A. L. Efros, Electronic 
Properties of Doped Semiconductors (Springer-Verlag, Berlin, 1984).
\bibitem{Bel86} B. I. Belevtsev, Yu. F. Komnik, A. V. Fomin, Fiz. 
Niz. Temp. 12 (1986) 821 [Sov. J. Low Temp. Phys. 12 (1986) 465].
\bibitem{Bel88} B. I. Belevtsev, Yu. F. Komnik, A. V. Fomin, 
Fiz. Tverd. Tela (Leningrad) 30 (1988) 2773.
\bibitem{Pakhom} A. B. Pakhomov, D. S. McLachlan, I. I. Oblakova, and A M.
Virnik, J. Phys.: Cond. Matter  5 (1993) 5313. 
\bibitem{Lee85} P. A. Lee and T. V. Ramakrishnan, Rev. Mod. Phys. 
57 (1985) 287.
\bibitem{Alt85} B. L. Altshuler and A. G. Aronov, in  
Electron-Electron Interactions in Disordered Systems, eds. A. L. Efros 
and M. Pollak (North-Holland, Amsterdam, 1985).
\bibitem{Dump} G. Dumpich and A. Carl, Phys. Rev. B43 (1991) 12074.
\bibitem{Gersh95} M. E. Gershenson, P. M. Echternach, and H. M. Bozler, 
Phys. Rev. Lett. 74 (1995) 446.
\bibitem{Bel95a} B. I. Belevtsev, E. Yu. Beliayev, V. V. Bobkov, and 
V. I. Glushko, Fiz. Niz. Temp. 21 (1995) 763 [Low Temp. Phys. 21 (1995) 
592].
\bibitem{Bel95b} B. I. Belevtsev, Yu. F. Komnik, and E. Yu. Beliayev,  
Fiz. Nizk. Temp. 21 (1995) 839 [Low Temp. Phys. 21 (1995) 646].
\bibitem{Buch84} A. V. Butenko, E. I. Bukhshtab, and V. V.
Pilipenko, Fiz. Nizk. Temp. 10 (1984) 773 [Sov. J. Low Temp. Phys. 10 
(1984) 407].
\bibitem{Most89} M. Mostefa, D. Bourbie, and G. Olivier, Physica 
B160 (1989) 186.
\bibitem{Nguen85} V. L. Nguen, B. Z. Spivak, and B. I. Shklovskii, 
Pis'ma v Zh. Eksp. Teor. Fiz. 41 (1985) 35 [JETP Letters 41 (1985) 42].
\bibitem{Shklov91} B. I. Shklovskii and B. Z. Spivak, in Hopping 
Transport in Solids, eds. M. Pollak and B. I. Shklovskii 
(Elsevier Science Publishers B. V., New York, 1991). 
\bibitem{Sivan} U. Sivan, O. Entin-Wohlman, and Y. Imry, Phys. Rev. Lett.
60 (1988) 1566.
\bibitem{Botg94} H. B\"{o}tger, V. V. Bryksin, F. Schulz, Phys. 
Rev. B49 (1994) 2447.
\bibitem{Eto} M. Eto, Phys. Rev. B48 (1993) 4933. 
\bibitem{Wang} X. R. Wang and X. C. Xie, Europhys. Lett. 26 (1997) 1111.
\end{thebibliography}
\end{document}